\documentclass[conference]{IEEEtran}
\ifCLASSINFOpdf
  \usepackage[pdftex]{graphicx}
\else
  \usepackage[dvips]{graphicx}
\fi
\usepackage[font=footnotesize, caption=false]{subfig}
\begin{document}
\title{The Distributed Network Processor:\\
a novel off-chip and on-chip interconnection network architecture}
\author{\IEEEauthorblockN{Andrea Biagioni,
Francesca Lo Cicero,
Alessandro Lonardo, 
Pier Stanislao Paolucci,
Mersia Perra,}
\IEEEauthorblockN{Davide Rossetti,
Carlo Sidore,
Francesco Simula,
Laura Tosoratto and
Piero Vicini}\\
\IEEEauthorblockA{INFN Roma c/o Physics Dep. ``Sapienza'' Univ. di Roma, p.le Aldo Moro, 2 - 00185 Roma, Italy\\ Email: [firstname.lastname]@roma1.infn.it}}


\maketitle

\begin{abstract}\boldmath
%Papers should clearly describe the nature of the work, explain its significance, highlight novel features, and describe its current status
One of the most demanding challenges for the designers of parallel computing architectures is
to deliver an efficient network infrastructure providing low latency, high bandwidth
communications while preserving scalability. 
Besides off-chip communications between processors, recent multi-tile (i.e. multi-core)
architectures face the challenge for an efficient on-chip
interconnection network between processor's tiles.
%% In the current scenario
%% of convergence between Multi-Processor System on Chip (MPSoC)
%% and High Performance Computing (HPC), we tackle this challenge designing a dedicated,
%% parametric, scalable network architecture for systems ranging from a single MPSoC to a massive HPC platform.
In this paper, we present a configurable and scalable architecture,
based on our Distributed Network Processor (DNP) IP Library, targeting
systems ranging from single MPSoCs to massive HPC platforms.

The DNP provides inter-tile services for both on-chip and off-chip
communications with a uniform RDMA style API, over a multi-dimensional
direct network --- see~\cite{Duato} for a definition of direct
networks --- with a (possibly) hybrid topology.
%
%% The DNP is a provider of communication services both at inter-tile
%% level, implementing an off-chip, inter-processor communication via RDMA
%% over a multi-dimensional network topology, and at intra-tile level, acting
%% as DMA controller for local memory access.
%
%
%% a)local    = intra-tile
%% b)remote   = inter-tile
%% c)on-chip  -> intra-tile & NoC
%% d)off-chip -> inter-tile su 3D
%% e)local  -> on-chip
%% f)remote -> off-chip & NoC
%% remote = off-chip || on-chip
%% local  = intra-tile 
%
%% Its main duty is to efficiently connect \textit{computational nodes}, defined as a single processing unit as well as a full-fledged MPSoC,
%% while acting as a DMA controller for \textit{local} communications, offloading the computational unit from this task.
%% Equipping each \textit{computational node} in a system with a DNP, its high level of
%% parametrization enables the design architect a considerable freedom in
%% implementing different network topologies.
%
It is designed as a parametric Intellectual Property Library easily
customizable to specific needs.
The currently available blocks implement wormhole, deadlock-free
packet-based communications with static routing.
%Where necessary virtual channels are DNP routing is guaranteed deadlock-free by the implementation of .
%DNP programmability is provided by an external $\mu$P through the DNP \textit{local slave interface}. 
%Furthermore the DNP is programmed via this interface,
%submitting commands in an asynchronous manner to a dedicated \textit{command queue} giving back results upon completion.

The DNP offers a configurable number $L$, $N$ and $M$ of ports ---
respectively intra-tile I/O ports to ensure connections among
elements within the same computational tile, on-chip communication
ones to link different tiles onto the same silicon die, and off-chip
communication inter-tile I/O ports to link those belonging to
different dies. ---
Because of the fully switched architecture, the DNP may sustain up to
$L+N+M$ packet transactions at the same time.

The DNP has been integrated into the design of an MPSoC dedicated to
both high performance audio/video processing and theoretical physics
applications.
We present the details of its architecture and show some promising
%results we obtained on the first current implementation.
%MP changed in
results we obtained on a first preliminary implementation.
 
%% In this customization, the $L=2$ \textit{local master ports} (needed
%% for the DNP acting as a DMA engine) are connected to the AMBA-AHB matrix via 
%% corresponding AMBA-AHB Master ports performing protocol adaptation.
%% The $M=6$ \textit{remote} ports are designed to support a 3D
%% toroidal lattice off-chip interconnection with a SerDes interface.
%% An additional \textit{remote} port ($N=1$) is used to connect the DNP
%% to the ST Spidergon NoC for on-chip communications.

%% Our future plans foresee to integrate the DNP in the next generation of Array Processor
%% Experiment (APE\footnote{http://apegate.roma1.infn.it/APE/}) machine, a
%% massively parallel super-computer, optimized for Theoretical Physics
%% and Bioinformatics simulations.

\end{abstract}



% \fi
\IEEEpeerreviewmaketitle

%------------------------------------------------------------------------------

\section{Introduction}
In the '70s, the ever-rising complexity of the numerical applications
employed by the scientific community started for the first time to
overcome the computing power offered by any single-processor computer
platform.
The need for higher computational capability together with both
scaling of silicon technology and power constraints, bolstered
investigation in the field of High Performance Computing (HPC).
From the beginning, the main issues related to HPC were scalability
and robustness, in particular for off-chip and off-board
communication.

A similar scenario is now unfolding in the embedded world.
The semiconductor industries, due to limitations similar to those the
HPC community faced in the past, are now all shifting their interests
from a monolithic processor design approach to Multi Processor System
on Chip (MPSoC)~\cite{Dally}~\cite{RAWtile}.
In this case, the challenge is to provide an efficient communication
infrastructure when traditional backbone bus systems fail on
guaranteeing full scalability and enough bandwidth between several
tens of on-chip processing units (micro-controllers, DSPs,
accelerators, etc).
Nowadays we are observing a convergence between the embedded and the
HPC systems, as the single HPC computational node becomes more and
more often a MPSoC.

In this scenario we propose a novel embeddable and fully synthesizable
hardware block, the Distributed Network Processor (DNP), that provides
a fully scalable and configurable network infrastructure (the DNP-Net)
for both on-chip and off-chip interconnection, suitable for platforms
scaling from a single MPSoC to a massive HPC system.
%a fully scalable parametric network infrastructure (the DNP-Net) for
%both on-chip and off-chip communication, suitable for platforms able
%to scale from a single MPSoC to a massive HPC system.
%
%Nessun block stands out for extraordinary innovation, Non siamo on the
%edge, ma l'insieme e` un buono playground per future innovative research.
%
%The scope of the DNP project is not to produce an fully developed
%end-product, rather to create a playground onto which future
%innovative research can be easily developed.
%
%While the scope of the DNP project is to pave the way for future
%innovative research that can be easily developed by extending current
%modules, it currently drives the networking part of real projects.

In the following, the elementary processing units connected by the DNP
are called \textit{tiles}, and multiple potentially-heterogeneous
tiles can be laid out on a single chip.  A single tile may be very
simple --- e.g. consisting of a single $\mu$P plus the DNP -- or quite
complex --- e.g. a $\mu$P, one or more DSPs plus the DNP, -- but every
one of them contains a DNP unit. The hierarchy may be further
deepened: multiple multi-tile chips may be assembled on a processing
board, and multiple processing boards plugged in a rack and wired
together to build a high-performance HPC parallel system.
The DNP is connected to the other devices inside the tile via a set of
so-called \textit{intra-tile} interfaces, while \textit{inter-tile}
interfaces are employed to connect a tile to the other ones, which by
the way may (on-chip) or may not (off-chip) reside onto the same chip.

A distinguishing feature of ours is that the same set of RDMA-like
communication primitives can be uniformly employed to address both
on-chip and off-chip tiles.
Traditionally RDMA-capable interconnects --- e.g. Infiniband, Myrinet
--- are quite expensive as of hardware resources, needing hundreds of
megabytes of memory, and high-speed offloading processors to properly
handle the complexities of the layered protocol stack or just the
virtual memory translations of standard operating systems (Linux,
Windows), running on commodity processors (x86, PowerPC, etc.).
On the one hand, typical HPC parallel numerical applications need full
control of the computing devices, badly tolerating multi-users
environments while being really allergic to swapping/paging regimes ---
i.e. they use no more than 99\% of physical system memory. ---
On the other hand, the data processing and real-time part of embedded
MPSoC applications need locked physical buffers used as targets of
DMA transfers to/from data acquisition devices,
% using either specialized memory allocators or custom low level software drivers
while the virtual memory support is really used only for the traditional part of
the application --- i.e. the user interface, configuration logic, etc.---
That is why most of the times the memory address translation features
--- e.g. virtual-to-physical mapping units or MMUs --- of commercial
processors, even if available, can be turned off, as is the case on
most embedded operating systems (vxWorks, uCLinux) or even IBM
BlueGene/L~\cite{IBMBGL}.
Having this in mind, the DNP has been optimized for the no-memory
address translation case, making it comparatively small in size and
relatively scarce in used silicon resources, opening the possibility to
put multiple DNP instances onto the same chip. As a consequence, the
RDMA primitives can be promoted from a low-level API, onto which
implementing Message Passing Interface (MPI) --- as is the case on
clusters with Infiniband, --- to a full-fledged system-wide
communication API, uniformly targeting both on-chip and off-chip
devices. In other words, the same RDMA API can be used throughout the full
hierarchy of devices.

The paper skips over all DNP software related topics (communication
libraries, simulators, etc..) which can be found
elsewhere~\cite{HDS}~\cite{DOL}. It is organized into four
sections. In the first one we give an overview of the DNP
architecture, describing its modular structure and the available
interfaces. In the second section we report on the use of the DNP in
the framework of the SHAPES project, which was actually the motivation
for its development in the first place. In the next one we collect
some results of our work and finally we write down our conclusions.

\section{DNP Architecture}

The DNP~\cite{DNP1} acts as an off-loading network engine to the
tile, performing both on-chip and off-chip transfers as well as
intra-tile data moving.
%AB
At the inter-tile level it operates as a network adapter, while at
intra-tile level it can be used as a basic DMA controller useful
for moving large data chunks --- i.e.\ between external DRAM and internal
static RAM. ---
%AB
%We simplified our design splitting inter-tile i

\begin{figure}[!t]
\centering
\includegraphics[width=3in]{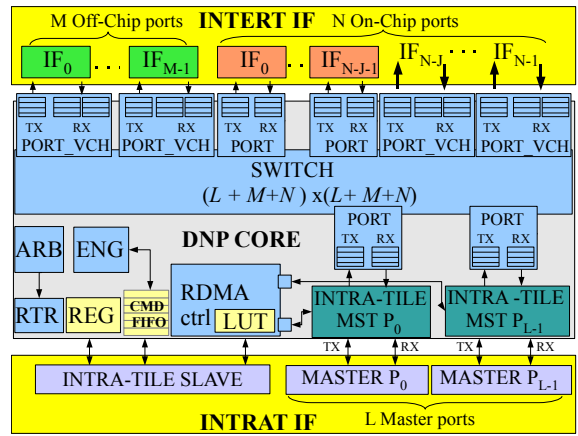}
\caption{A block diagram of the DNP architecture: the modular design
  separates the architectural core (\emph{DNP CORE}) from the
  interface blocks (\emph{INTRAT IF} and \emph{INTERT IF}). The
  inter-tile interfaces can be either off-chip or on-chip, and in
  either cases they can be equipped with virtual-channels. The
  intra-tile slave interface is basically used by the software to
  configure and program the DNP. The intra-tile master interfaces are
  controlled by the DNP and are used to stream data in and out. }
\label{fig_DNP_arch}
\end{figure}

The DNP has been developed as a \emph{parametric} Intellectual Property
library, meaning that some fundamental architectural features can be
chosen among the available ones, others can be customized at
design-time and new ones can be developed.
A highly modular design has been employed, separating the
architectural core (\emph{DNP CORE}) from the interface blocks
(\emph{INTRAT IF} and \emph{INTERT IF}), as shown in \figurename
\ref{fig_DNP_arch}.  The Core and the
Interfaces are connected by a customizable number of \textit{ports}.

% that in the DNP library may be choose between simple ports or Virtual CHannel (VCH) equipped ports.

%The DNP is based on a switch system

%RICONFIGURABILITA'

\begin{figure}[!t]
\centering
\includegraphics[width=3in]{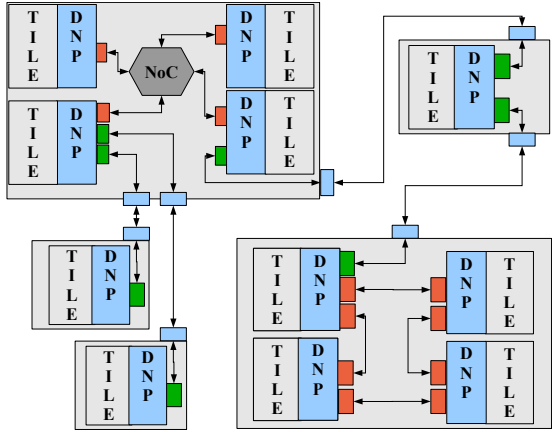}
\caption{Examples of on-chip and off-chip network topologies and
  services offered by reconfiguring the parametric DNP.}
\label{fig_DNP_net}
\end{figure}

This flexibility allows tailoring the DNP architecture to the
particular use-case, choosing the topology of the DNP-Net:
it is just needed to define the number $M$ of inter-tile off-chip
ports, the number $N$ of inter-tile on-chip ports and the routing
algorithm to automatically achieve the desired configuration (\figurename \ref{fig_DNP_net}).

For example, in the SHAPES project~\cite{SHAPES1}, as we will show in
section \ref{SHAPEScasestudy}, our IP library has been used to implement
a 3D torus network topology in a multi-tile system architecture.
%
%To complete the customizability options of the DNP, there is also the chance to choose
%the number $L$ of master ports used to read/write data into the source/destination device
%inside the tile. 
%
Besides to fit with the required network I/O bandwidth, a suitable
number $L$ of intra-tile master ports can be chosen.
%level, it manages inter-processor communication via RDMA, implementing a multi-dimensional
%network topology; at intra-tile level, it acts as DMA controller for local memory access.
%\textbf{Implementabili fino a 3 slave}

%The DNP has been developed as \emph{parametric} Intellectual Property
%library, meaning that many fundamental architectural features can be customized at
%design-time.
The DNP architecture is a crossbar switch with configurable routing
capabilities operating on packets with variable sized payload. The
implementation of virtual channels~\cite{DALLY2} on incoming
switch ports guarantees deadlock-avoidance.

The DNP gathers data coming from the intra-tile ports, fragmenting the
data stream into packets (\figurename \ref{fig_DNP_pkt}) which are
forwarded to the destination via the intra-tile or inter-tile ports
depending on the requested operation.

%To instrument the DNP to transfer data from one point to another into the system, a command must be pushed into its local slave port.
%The command contains basic information as source and destination node, length of the data burst to be moved, 
%which local port must be used to perform local transactions 
%and other control information that will stand over the transaction.
%Once the command is issued

\subsection{RDMA architecture and communication robustness}
\label{RDMA}

%Possibilità di proteggere le zone di memoria. 
%i target dell'operazione RDMA devono esser pubblicati e sono nella LUT.
%Il concetto della completion
%event based API
%in hardware abbiamo PUT e GET 
%
%The RDMA LUT can be written and read by the application and contains
%the list of buffers that xhave been published by the application.
%More details will be found on the section (see sect. \ref{RDMA}).

The DNP RDMA features may be condensed into three concepts:
\begin{itemize}
\item A Command Queue, implemented as a hardware FIFO queue (\emph{CMD FIFO}),
  where the software pushes RDMA commands to be asynchronously
  executed by the DNP. A DNP command is composed by seven words
  containing information necessary to perform the required data
  transport operation.
\item A Completion Queue (\emph{CQ}), which lives in the tile memory
  and is treated as a ring buffer, where the DNP writes events, which
  are simple data structures, and software reads them. Events are
  generated as commands are executed and incoming packets are processed.
\item A Look-up Table (\emph{LUT}), a hardware memory block embedded in the
  DNP which is accessible by software through an intra-tile interface.
\end{itemize}

\begin{figure}[!t]
\centering
\includegraphics[width=3in]{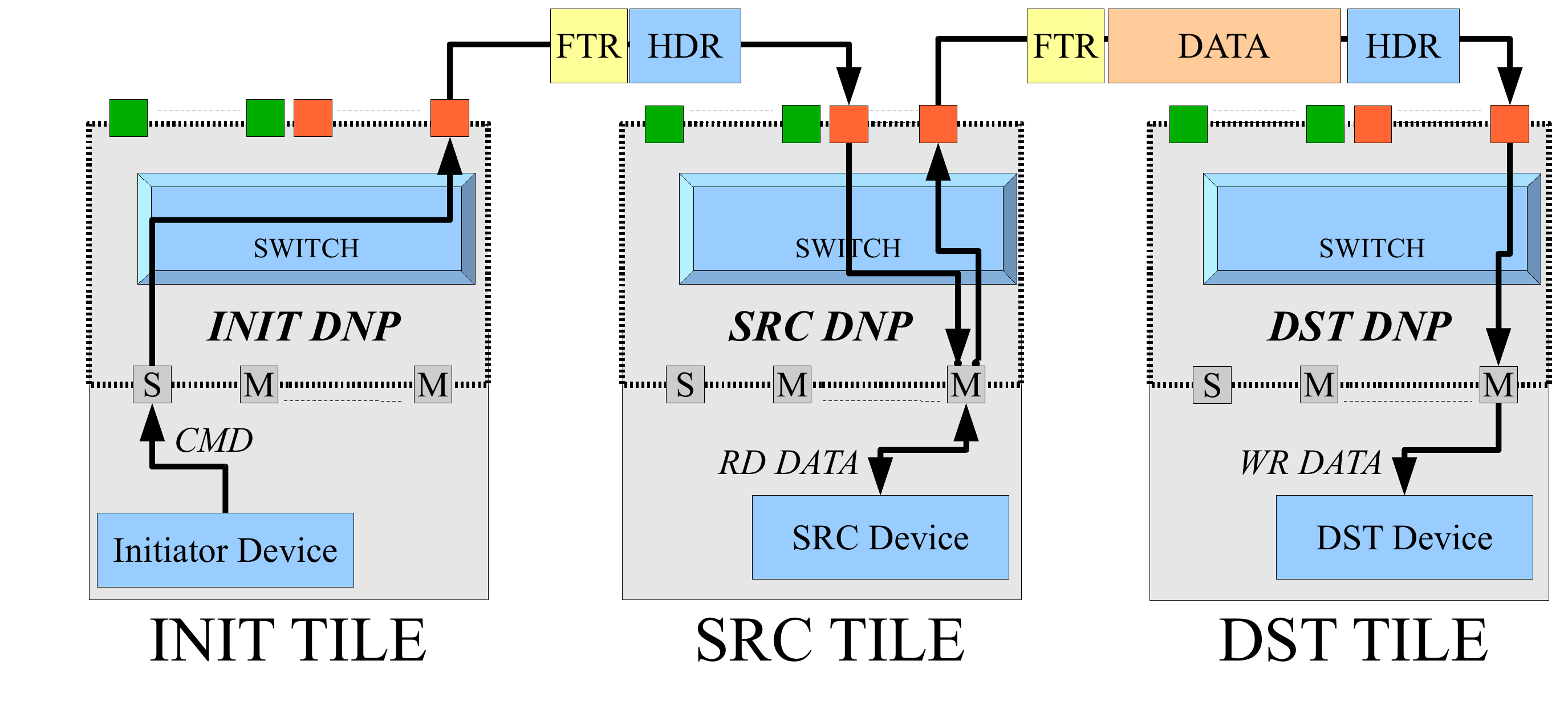}
\caption{Data transfer example: a RDMA GET operation characterized by different initiator (\emph{INIT DNP}), source (\emph{SRC DNP})
and destination (\emph{DST DNP}). The most common use is with INIT $==$ DST.}
\label{fig_DNP_GET}
\end{figure}

The supported RDMA commands are described by a compact set of
parameters: the command code (\textbf{LOOPBACK}, \textbf{PUT},
\textbf{SEND} and \textbf{GET}), the source memory address and DNP
(\emph{SRC DNP}), the destination memory address and DNP (\emph{DST
  DNP}), the length in words.

The \textbf{LOOPBACK} command describes a memory move operation in
which the content of a (local) memory area is copied into another
one. One intra-tile interface is used to fetch data and another one to
write it to the final destination.
The RDMA \textbf{PUT} and \textbf{SEND} commands generate a one-way
network transaction; the DNP executes them by reading the local buffer
and produces a packet stream targeted to the destination DNP.
The RDMA \textbf{GET} commands is a two-way transaction in which the initiator DNP (\emph{INIT DNP})
generates a \textit{request} packet destined to the source DNP, which
in its turn will generate a data packet stream toward the destination
DNP (see \figurename\ref{fig_DNP_GET} for the most general,
three-actors situation).
Most of the times the destination DNP is the same as the initiator so
it collapses to the usual RDMA GET operation as found in other
networks.
After execution of each command, the DNP optionally writes an event in
the \emph{CQ}, so that the software may acknowledge the command as
executed, freeing the source data buffer for further use.

The buffers which are used as a source in a RDMA command need no
special care while those which are destination have to be
pre-registered into the \emph{LUT} by the software.
The \emph{LUT} is organized in records, each one containing the buffer
physical start address, length and some flags.
When a packet is received, the \emph{LUT} is scanned in search for an
entry matching the packet destination buffer; only in this case the
operation is carried on.
After writing (\textbf{PUT} and \textbf{SEND}) or reading (\textbf{GET})
the data, a corresponding event is written in the \emph{CQ}, so that
the software may carry on further operations --- e.g.\ deregistering
the buffer, sending back some acknowledge packets, etc. ---
\textbf{SEND} packets are similar to \textbf{PUT} ones with null
destination address, so that the first suitable buffer in the
\emph{LUT} is picked up and used as the target buffer. They are vital
for the software application to bootstrap the RDMA protocol at the
beginning --- e.g.\ after buffers allocation to communicate their
address to the other DNPs, --- in the \textit{eager} communication
protocol, and more in general to exchange small amounts of data ---
e.g.\ protocol packets. --- On the other hand, \textbf{PUT} packets
are mostly intended for the \textit{rendezvous} protocol.

\subsection{DNP packets}
\label{Packets}

\begin{figure}[!t]
\centering
\includegraphics[width=3in]{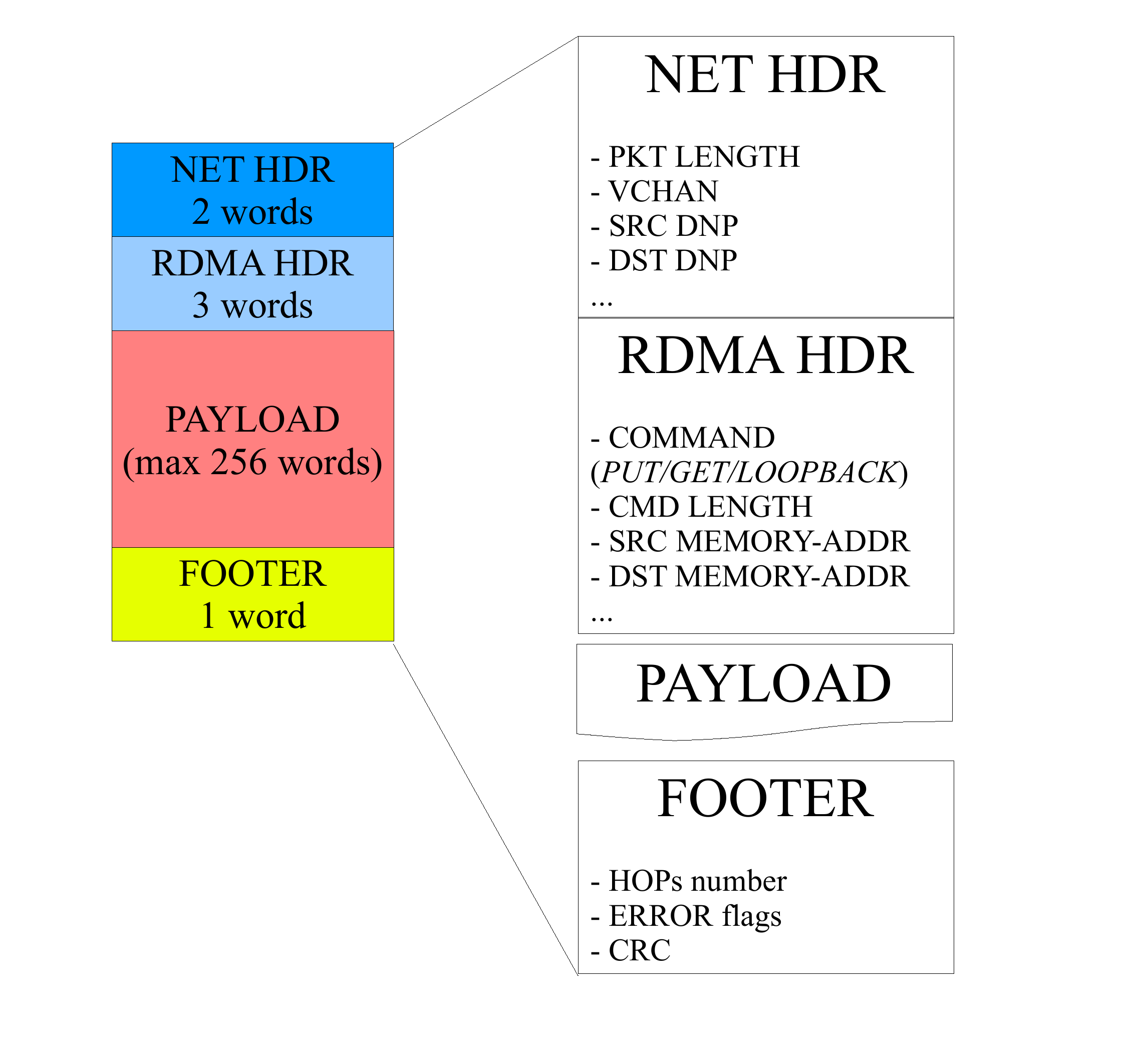}
\caption{DNP packet overview: the packet is made of fixed size header
  and footer and variable number of payload words. The payload can be
  up to 256 words. There is an optional space for an integrity check
  code (CRC) and corrupted packets are flagged by a single bit in the
  footer.}
\label{fig_DNP_pkt}
\end{figure}

A data sending operation may generate one or more outgoing
packets. The DNP hosts a hardware fragmenter block which automatically
cuts a data words stream into multiple packets stream. A packets is
made up of a fixed size envelop plus a variable word sized payload
(\figurename\ref{fig_DNP_pkt}). In the current implementation, the
header is split into a network (\emph{NET HDR}), which basically
carries routing information, and a RDMA header (\emph{RDMA HDR}),
which is processed only by the destination DNP; the footer hosts the
packet integrity code and the corruption flag.

Every DNP is uniquely addressed by a 18 bit string, whose
interpretation depends on the exact details of the network topology;
address decoding is done in the router module and must be customized
accordingly. For instance, in a 3D Torus network those bits can be
evenly split into a $(x,y,z)$ triplet, while on a NoC based design
there could be an additional internal coordinate, i.e.\ a 4-tuple like
$(x,y,z,w)$.

\subsection{Reliability guarantees}
\label{Guarantees}

The DNP architecture lives on some basic assumptions:
\begin{itemize}
\item Neither the destination DNP nor the in-between transient DNPs
  are allowed to drop packets.
\item Each packet is always reliably transferred to its destination.
\item Packet corruption is only allowed in the payload and has to be
  detected and marked in the footer, so that it can be handled by the
  application.
\item There is no low level protocol to signal packet delivery,
  e.g.\ by n/ack packets. If necessary, it can be implemented by the
  application.
\end{itemize}
These assumptions turn into requirements for the hardware
architecture; it has to be robust enough to protect at least the
packet envelop (header and footer), e.g.\ to avoid bad routing due to
corrupted headers.
On-chip bit error rate (BER) is assumed negligible, at least at the
single tile level with mature silicon processes. Most of the current
on-chip inter-tile communication technologies (NoC) claim to control
the BER on their own. So the inter-tile off-chip interfaces are the
one mostly hit by the above requirements: they have to employ suitable
flow-control techniques to avoid packet dropping; some form of
redundancy has to be employed to protect the packet envelop from
corruption, e.g.\ using checking and retransmission techniques.

\subsection{Core}
\label{Core}

The modular approach used to design the DNP architecture is also
applied to the DNP core (\figurename\ref{fig_DNP_arch}), therefore
its main functionality is split in several separated modules.

%La parte di gestione dei comandi RDMA è gestita dal RDMA_BLOCK. L'Engine (o intratile) e l'RDMA BLOCK sono 2-3 macchine a strate sparse che però implementano la stessa funzione.

Devices inside the tile require DNP primitives issuing
\textit{commands} via the the intra-tile slave interface
(\emph{INTRA-TILE SLAVE}).
The Engine (\emph{ENG}) fetches commands from the \emph{CMD FIFO} and
uses them to fill out the packet header.
The payload data are read by an intra-tile transaction using
information in the RDMA Controller block (\emph{RDMA ctrl}) and the
newly created packets are forwarded through the Switch port towards
the proper inter-tile ports \emph{PORT\_VCH} or \emph{PORT}.
%The need of implementing virtual channel depends on the DNP being
%responsible for assuring deadlock avoidance.

The routing logic (\emph{RTR}) configures the \emph{SWITCH} paths
between the DNP ports, sustaining up to $L+M+N$ simultaneous packet
transactions.
If more than one packet requires the same port, the arbiter block
(\emph{ARB}) applies the arbitration policy to solve the contention.
The parametrized number $L$ of intra-tile master ports
(\emph{INTRA-TILE MST}) guarantees the connection among elements within
the same tile.

%Packets from/to the Rx/Tx ports are translated by this block in a proprietary \textit{DNP bus protocol} based on address, data and control signals. 
On receiving of a packet, an intra-tile transaction is carried out
with information from the \emph{RDMA ctrl} block, which wraps the
\emph{LUT} inside.
%The intra-tile master is also in charge of inquiring the \emph{RDMA ctrl} block. 
Each RDMA transaction is followed by a \textit{completion} operation
(see sect. \ref{RDMA}).

Besides the \emph{CMD FIFO}, both a set of registers (\emph{REG}) and
the RDMA Look-Up Table (\emph{LUT}) are accessible through the
intra-tile slave port.
The registers are used to expose status information and to configure
the DNP functionality; i.e.\ hand-shake protocols among blocks are
often time-out based with exception rising, so that time-out
thresholds, as well as arbitration logic choice and the port priority scheme, are
configurable this way.
Moreover, some registers allow for resetting and dis/enabling of
blocks inside the DNP at run time by software.

\begin{figure}[!t]
\centering
\includegraphics[width=3in]{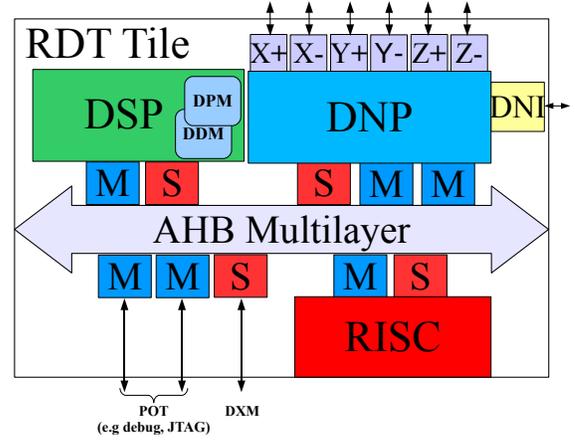}
\caption{The SHAPES RISC+DSP+DNP Tile (RDT) is an example of
  integration of the DNP in a commercial environment (Atmel Diopsis
  940) to provide networking capabilities. In this drawing, the
  \emph{M} and \emph{S} boxes are respectively \emph{master} and
  \emph{slave} interfaces. DPM and DDM are DSP Program and Data
  memories. POT is the set of peripherals. DXM is the external data
  memory controller.}
\label{fig_RDT}
\end{figure}

\subsection{Intra-tile and inter-tile Interfaces}

%As previously  described, the DNP is designed with high level of modularity.
As previously described, the modularity depicts the DNP design.
The DNP may be adapted to different environments with minor rewriting
efforts, being the interfaces completely independent from the core.
Given network topologies lead to different protocol choices as well as
error checking and correction requirements.
%Keeping these features indepents from the Core allows making the most
%out of the DNP features, enhancing them with the specific design
%implementation.
Keeping some of these features out of the Core module makes it smaller
and cleaner, moving the potential complexity to the inter-tile or
intra-tile interfaces.
%Therefore the implementation of these communication blocks è totalmente  lasciando tipo in questo modo la possibilità di utilizzare tutte le 
%funzionalità del dnp-core

%Different topology may need different approach on error checking and correction. 
At intra-tile level the DNP exposes a proprietary bus protocol 
while at inter-tile ports a FIFO like signaling is used. 

The intra-tile interfaces are in charge of translating the DNP
transactions into the particular protocol used inside the tile. These
interfaces are specific to the different bus architectures ---
e.g.\ AMBA-AHB, AMBA-AXI, PCI, PowerPC, etc. ---
In our library we provide AMBA-AHB adaptors for intra-tile
communication, either for the master (\emph{MASTER}) and the slave
(\emph{SLAVE}) interfaces.
On the other hand, PCI/PCI-X/PCI-e commercial cores can be easily
used by developing customized wrapper blocks.
The implemented \emph{INTRA-TILE SLAVE} adapter maps the registers,
the LUT and the command queue.
%However is possible to have up to three \emph{INTRA-TILE SLAVE}, one
%for each of the mantioned above blocks.

The inter-tile interfaces are responsible for on-chip and off-chip
interconnection.
While the DNP can be connected to highly specialized on-chip network
infrastructure --- e.g.\ a NoC, ---
%the communication may take advantage of NoC like solutions. 
however the DNP infrastructure allows for point-to-point
interconnection with simple interface blocks.

A very wide range of solutions may be applied for off-chip
communication depending on the requirements.
%Wired or wireless solutions may require quite differnt
%implementation.
In our case study (see sect. \ref{SHAPEScasestudy}), the
on-chip communication is NoC based while for off-chip we provide a
bidirectional Serializer/Deserializer (Ser/Des) with error check,
DC-balance and re-transmission capability.
The NoC used is the ST-Spidergon, while the Ser/Des block was a
proprietary solution.

%------------------------------------------------------------------------------

\begin{figure}[!t]
\centering
\includegraphics[width=3in]{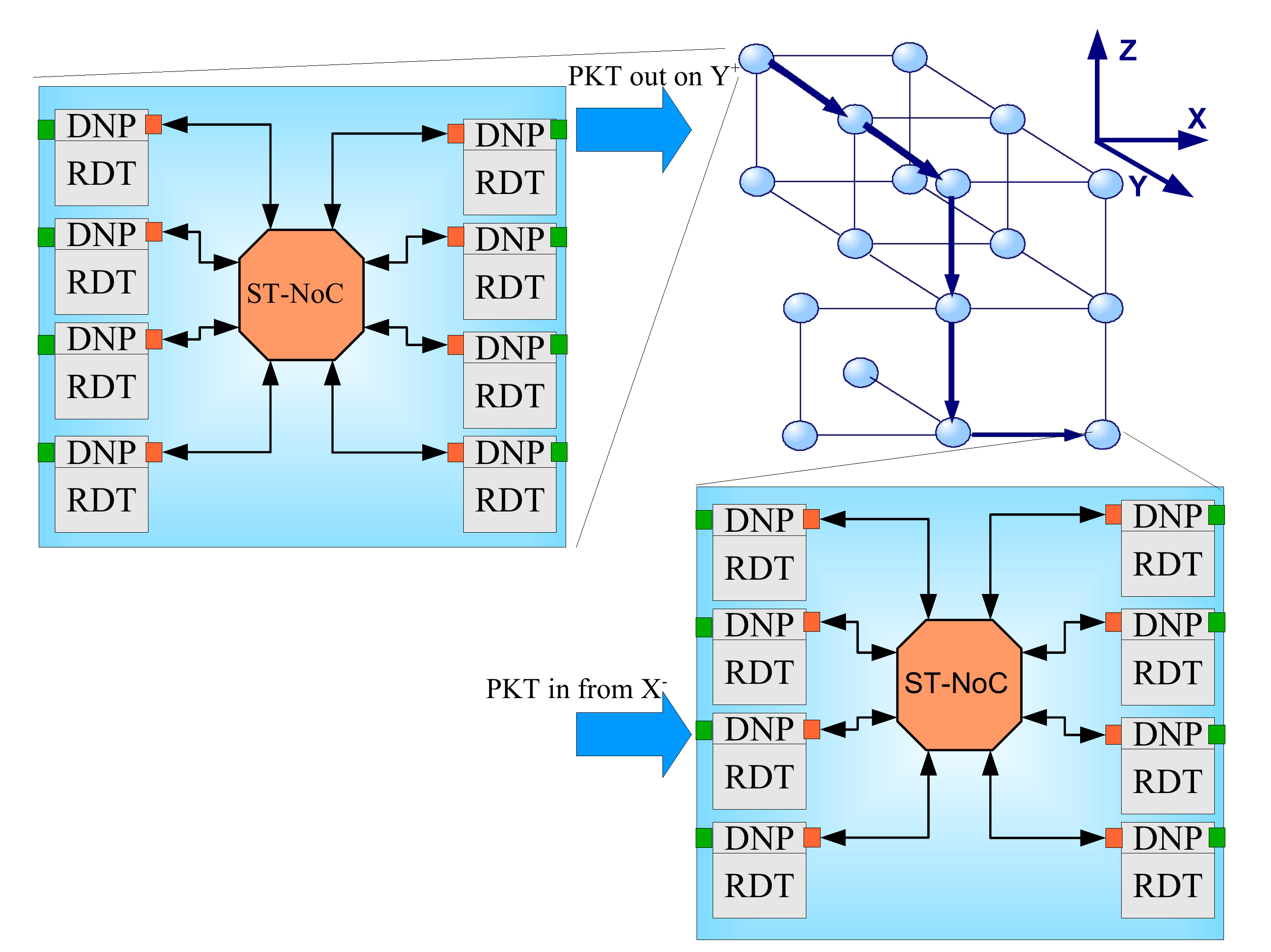}
\caption{8-RDTs SHAPES chip arranged in a 3D lattice. In each chip the 8 RDT tiles are connected by the ST-Spidergon NoC.}
\label{fig_RDTchipInLattice}
\end{figure}

\section{A Case Study: the SHAPES computing architecture}
\label{SHAPEScasestudy}

The DNP was developed in the context of the SHAPES European project, proposing a scalable HW/SW design style
for future CMOS technologies.

Besides the DNP, a typical SHAPES tile contains a VLIW floating-point
\emph{DSP} (Atmel mAgicV) with its on-chip Distributed Data/Program
Memory (\emph{DDM}-\emph{DPM}), a \emph{RISC} (ARM9), an interface to
Distributed External Memory (\emph{DXM}) and a set of peripherals
called the Peripherals on Tile (\emph{POT}) e.g. JTAG, Ethernet, etc.;
in SHAPES jargon, this tile configuration is called RISC+DSP+DNP Tile
(\textit{RDT}) (\figurename \ref{fig_RDT}).

The SHAPES network connects on-chip and off-chip tiles, weaving a distributed packet-switching network.
%Within the SHAPES tile, the DNP was developed to provide data transport functionalities to the tile, performing inter-tile,
%both on-chip and off-chip, and intra-tile communication.

The SHAPES software tool-chain provides a simple and efficient
programming environment for tiled architectures. SHAPES proposes a
layered system software~\cite{SHAPES1}, which dramatically helps the programmer to
harness the intrinsic complexity of fully scalable multi-processor
systems.

\subsection{The DNP in the SHAPES architecture}

In the SHAPES framework the intra-tile interfaces are AMBA-AHB
standard connected to the DSP and RISC through an AMBA-AHB multilayer
bus (\figurename \ref{fig_RDT}).
%We map the LUT, registers and cmd\_fifo on one single AHB Slave while
%two AHB Master ports are used to read/write data into the
%source/destination device.
For on-chip communications, the SHAPES architecture avails itself of
the ST-Spidergon Network-on-Chip~\cite{NoC1}~\cite{NoC2}~\cite{NoC3}
to which the DNP is connected by a specific interface, called the DNP
Network-on-Chip Interface (\emph{DNI}).
The 3D Torus topology has been adopted for off-chip networking
(\figurename \ref{fig_RDTchipInLattice}), with all node-connecting
bidirectional links, which needs a total of six inter-tile interfaces
per DNP.
This configuration of ports gives the particular SHAPES tile a render
of the DNP with $L=2$, $M=6$ and $N=1$.

%The DNP acts as a DMA controller (e.g. it moves data to and from the
%DSP internal data memory DDM and the tile memory DXM) when source and
%destination devices both belong to the same tile.
%
%When the packets reach the destination DNP, an intra-tile transaction
%towards the destination device is performed via the AHB Master port.
%
%To start a data transfer, a command must be issued on the DNP.  The
%DNP receives commands from the RISC or DSP processors on the AMBA AHB
%slave port.

The DNP applies a deterministic routing policy to implement
communications on the 3D torus network.
The coordinates evaluation order (e.g first Z is consumed, then Y and
eventually X ...)  can be chosen at run-time by writing into a
specialized priority register; in this way, the routing scheme is
configurable to a certain extent.
%Whenever the destination DNP is reached an intra-tile transaction
%towards the destination device is executed via the AHB Master port,
%otherwise an inter-tile communication is performed.
%
%This configuration of ports gives the particular SHAPES tile a render of the DNP with $L=2$, $M=6$ and $N=1$.
%The DNP implements 2 Virtual Channels so that routing is guaranteed to
%be deadlock-free on the 3D links.
%A set of registers are exposed through the AHB Slave port to configure, control and read the DNP status.
%The DNP supports Remote Data Memory Access (RDMA) programming model and it is ready to also support
%Systolic Communications and Message Passing Interface (MPI).

\subsubsection{The On-Chip Interface}

the DNI is the on-chip bidirectional interface handling DNP
transmissions to/from the ST-Spidergon NoC.
The communication protocol implied is a hand-shake protocol based on a
request/grant policy.
This interface includes a sub-module that verifies data by means of a
Cyclic Redundancy Check (CRC).
During the packet delivery process a CRC is computed and transmitted
together with the footer.
On receiving, that CRC is recalculated and checked, so in case of
transmission errors a bit in the footer is set and the packet goes on
its way.
The software can detect packet corruptions by analysing the footer.
%
%The footer is one of the completion word therefore, reading it, the software is being aware of the error occurred.

The ST-Spidergon NoC implements deadlock avoidance by its own,
therefore no virtual channels are necessary on the DNP port side.

\subsubsection{The Off-Chip Interface}

%The DNP implements 2 Virtual Channels so that routing is guaranteed to
%be deadlock-free on the 3D links.
%
%% As for the NoC Block Interface and the other inter-tile blocks, the Inter-tile interface are implemented by two separated blocks: the protocol-dependent and the protocol-independent blocks.  
%% The protocol independent block is a logic block which contains two FSMs. (one for the reception of data RXFSM and one for the transmission TXFSM) with the same features described so far for the Rx and Tx FSM (push and pop packets on the FIFOs), 
%% whereas the protocol-dependent blocks are the serializers and de-serializer blocks with CRC error check and DC-Balance feature.

%% In this section we will show functionalities and implementation of the SerDes block, the core of the
%% 3D Torus physical link Controller (3DTC). Each block is responsible for one spatial dimension
%% (X+, X-, Y+, Y-, Z+, Z-).

the inter-tile off-chip interface has a parallel clock SerDes
architecture, employing Double Data Rate signaling in order to reduce
packet transmission latency.
Special encoding and a DC-balance block guarantee the quality of the transmission line.
%
%This interface enumerates how many $1$ and $0$ bits travel along the
%physical link, in order to get a time-averaged null current. 
%
%The balancing is performed inverting the word.
The balancing is performed inverting the transmitted word to equalize
the number of $1$ and $0$ bits in time.
Furthermore, the inter-tile off-chip interface implements the
mesochronous clocking technique in order to handle the clock-phase
skew between communicating DNPs.

It manages the data flow encapsulating the DNP packets into a light,
low-level protocol able to detect transmission errors via CRC, and
includes a memory buffer to re-transmit the header and the footer in
case of transmission errors.
Therefore the protocol assures the delivery of the packet, avoiding
non-recoverable situations where badly corrupted packets (with errors
in the header or footer) pose threat to the global routing.
%
%In order to reduce complexity of the interface (in terms of
%size of required buffers to operate re-transmission) and transmission latency,
%the checksum of the payload is only checked but not re-transmitted.
%
In case of packets with payload errors (signaled by the footer), the
software communication library is in charge of handling the situation.

The chosen CRC polynomial generator is the same used for the inter-tile on-chip
interface: it is the industry-standard, well-known CRC-16.
% ($x^{16} + x^{15} + x^2 + 1$).
Past experiences collected on a custom developed massively parallel
computing architecture~\cite{apeNEXT}\cite{APENET} has given us insight into the
bit error rate of a typical off-chip LVDS link, while proving the
CRC-16 as a satisfactory solution.
% Supported by
%this experience and assuming the DNP to be employed with a similar
%off-chip link type, the same CRC was deemed sufficient to obtain an
%equivalent level of protection for the transmitted data.
%
%% In this particular trial implementation, each 3DTC has a mux/demux ratio equal to 16:1 and it is
%% connected to the adjacent replicas via three wires: two data lines and one clock line. We also
%% designed a slightly different version, where mux/demux ratio is 4:1 (8+1 lines per channel), to
%% demonstrate the scalability of the architecture.
%% The DNP protocol is packet-based. The packet is composed of 32-bit words and its structure is
%% composed of five header words, the payload with variable length, and one footer word.
%
%A register holds the result of the comparison of the received CRC with the computed CRC (ACK/NACK).
%Redundancy provides a greater level of protection against transmission errors.
%
%The inter-tile off-chip interface manages the data flow preventing the FIFO overflow, providing a
%signal to stop the data transmission when the free space inside a FIFO becomes smaller than a configurable
%threshold.

\subsection{Exploring the DNP high configurability}
\label{Exploring}

\begin{figure*}[!t]
\centerline{
\subfloat[On-chip communication ST-NoC based (MTNoC)]{\includegraphics[width=3.5in]{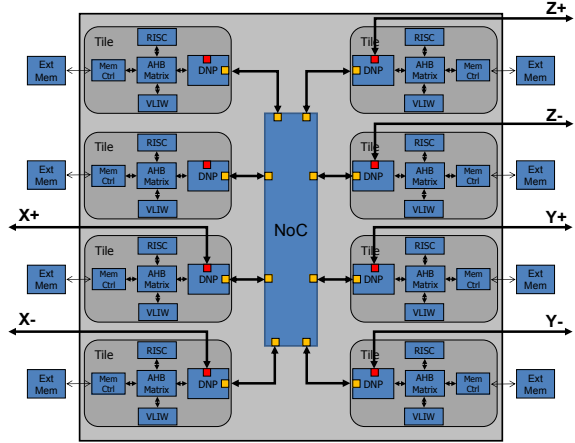}%
\label{fig_onChip_NoC}}
\hfil
\subfloat[On-chip communication 2D-DNP based (MT2D)]{\includegraphics[width=3.5in]{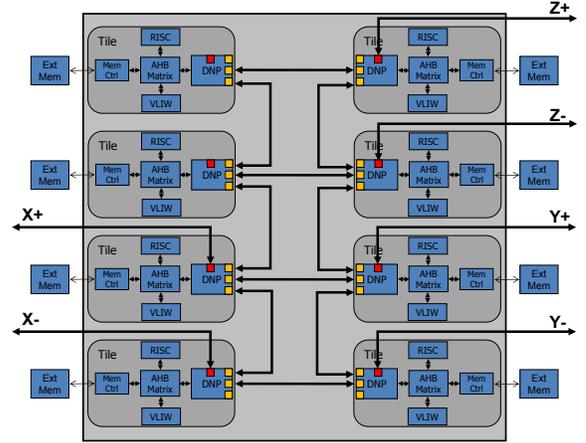}%
\label{fig_onChip_interTile_ports}}}
\caption{Details of the explored solutions: MTNoC and MT2D}
\label{fig_P2PvsNoC}
\end{figure*}

Within the SHAPES project, thanks to the high level of parametrization
offered by the DNP, we were able to propose different solutions for
the inter-tile on-chip network.
Besides the ST-Spidergon NoC (MTNoC), an alternative solution has been
proposed that involves an on-chip $2D$ mesh network, obtained using
the DNP inter-tile ports connected point-to-point (MT2D).

This experiment allowed having two solutions suitable for possibly
different application requirements
(\figurename\ref{fig_P2PvsNoC}).

%\begin{figure*}[!t]
%\centerline{\subfloat[Case I]\includegraphics[width=2.5in]{subfigcase1}%
%\hfil
%\subfloat[Case II]{\includegraphics[width=2.5in]{subfigcase2}%
%\label{fig_second_case}}}
%\caption{Simulation results}
%\label{fig_sim}
%\end{figure*}
%------------------------------------------------------------------------------

\section{Results}

The DNP architecture has been validated in the SHAPES project by using
the DNP TLM SystemC~\cite{Huang08} model integrated in different
multi-tile processor configurations.
In particular, the DNP was employed in benchmarking the SHAPES
architecture on a kernel code~\cite{QCD} for Lattice Quantum Chromo
Dynamics (LQCD), and tested on a system configuration of 8 RDTs
arranged in a $2 \times 2 \times 2$ 3D topology (one of those depicted in
\figurename\ref{fig_P2PvsNoC}).
%
%% The LQCD kernel is the fundamental set of operations at the base of
%% the most complex and computation-intensive applications used in this
%% branch of theoretical physics (which is a primary interest of INFN);
%% in this application, the DNP is a valuable asset. 
%% Due to the parallelism inherent to the LQCD, the DNP becomes an
%% independent agent that, offloading the cumbersome and time-consuming
%% task of moving data among tiles, lets the computational part show its
%% full potential, endowing the system with the necessary scalability and
%% bandwidth.

\begin{figure}[!t]
\centering
\includegraphics[width=2in]{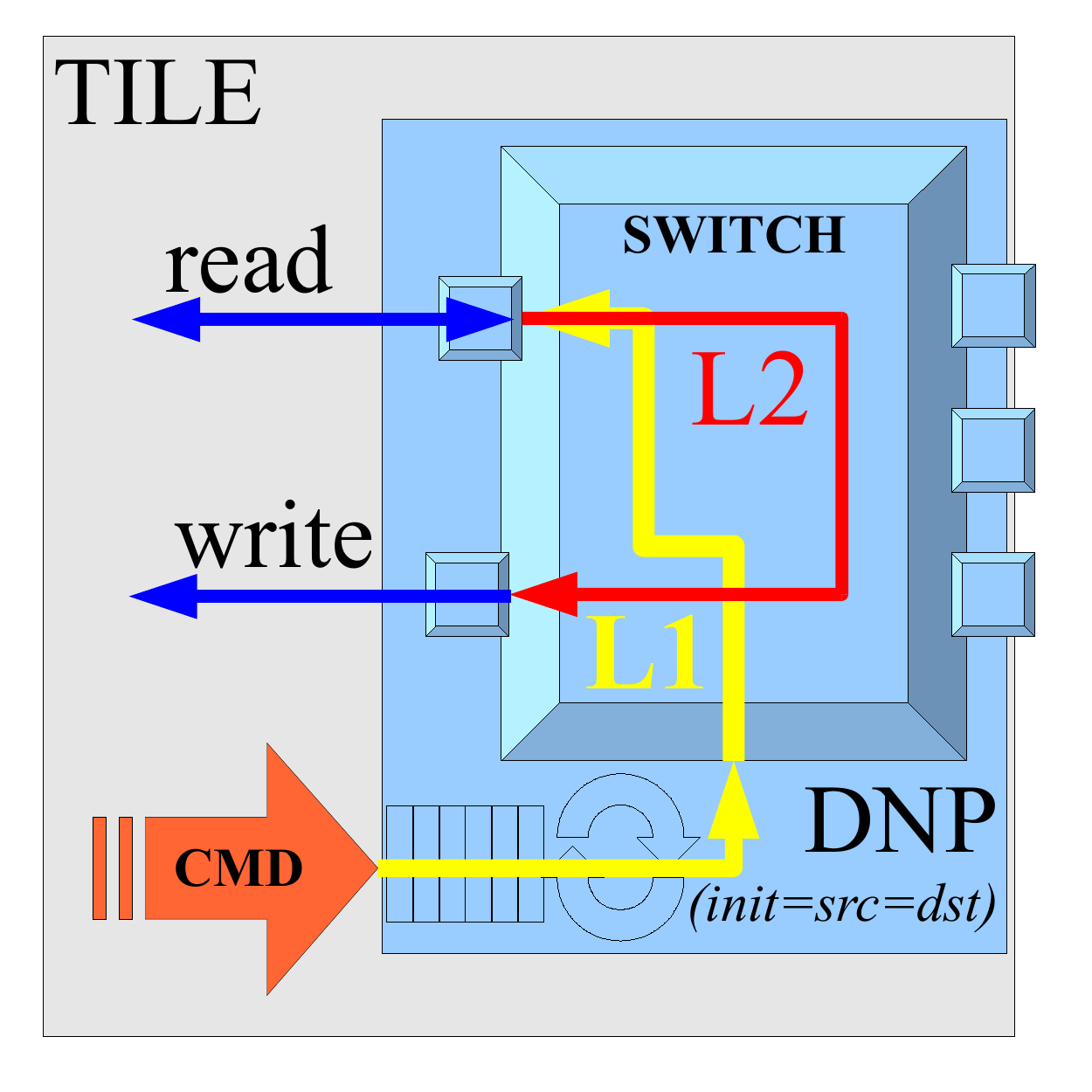}
\caption{Details on the timing measurement solution adopted for a
  \textbf{LOOPBACK} operation. $L_1$ represents the time elapsed
  between the command issuing and the beginning of the read intra-tile
  transaction, while $L_2$ is the time necessary to complete
  \textbf{LOOPBACK} operation and to begin the write intra-tile
  operation.}
\label{LOOPBACK_meas}
\end{figure}

At the same time, an RTL model (written in VHDL) has been used to
simulate and validate the DNP implementation, with both ad-hoc stimuli
test vectors and mixed SystemC-VHDL simulations.
%with realistic C++ applications generating stimuli for the model.
%
Along with preliminary measures of intra-tile and inter-tile bandwidth
and latency, we estimate area consumption and power dissipation on a
early place-and-route layout. In the following, bandwidth figures are
meant unidirectional, i.e.\ for each direction, unless explicitly
specified.
%, considering that anyway all interfaces are fully bidirectional.

Depending on the particular intra-tile interface adopted in the
destination design --- e.g. matching a bus with reduced width or with
lower frequency, --- the DNP intra-tile port is able to sustain up to
1word/cycle (1 word equals 32 bits), so the resulting intra-tile
bandwidth is $BW_{int} = L \times 32 = 64 \mbox{bit/cycle}$, roughly
4GB/s at 500MHz, or $4+4\mbox{GB/s}$ bidirectional bandwidth.
The latency for intra-tile communications --- i.e.\ the time interval
elapsed from the LOOPBACK command reaching the \emph{CMD FIFO} to the
first outgoing data surfacing onto the bus, as in \figurename\ref{LOOPBACK_meas} ---
 is $L_{int}= L_1+L_2 \simeq 100$ cycles,
equal to 200ns at the target frequency.
% while of course the
%bandwidth is half the peak, as both ports are used for the same data
%stream.

\begin{figure}[!t]
\centering
\includegraphics[width=3in]{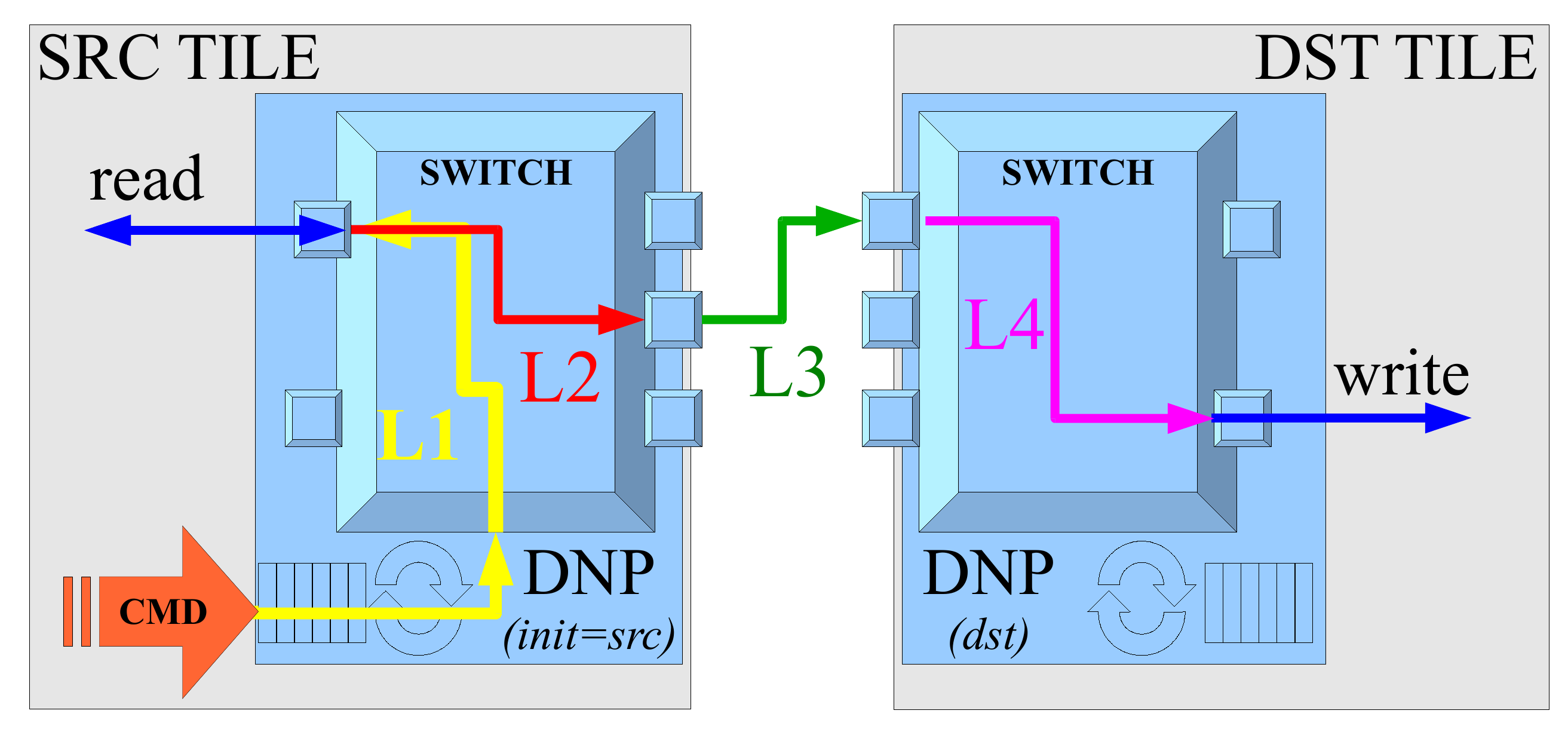}
\caption{Single-hop \textbf{PUT} command with 1 word payload. $L_1$
  represents the time elapsed between the command issuing and the
  beginning of the read intra-tile transaction; $L_2$ is the time
  necessary to transmit the first header word of the packet to the
  appropriate inter-tile interface across the switch; $L_3$ is the
  transmission time over the serialized off-chip interface toward the
  DST~TILE; $L_4$ is the time down to the intra-tile write
  operation. }
\label{PUT_meas}
\end{figure}

\begin{figure}[!t]
\centering
\includegraphics[width=3in]{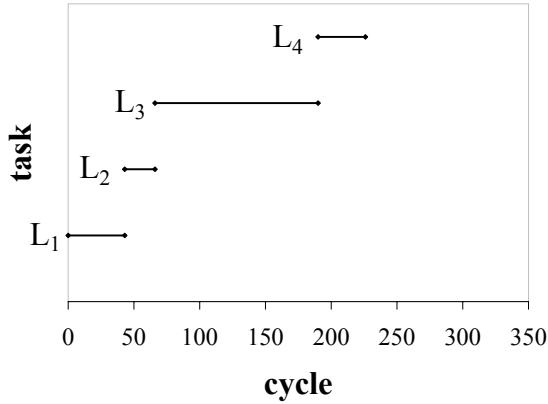}
\caption{The broken-out timings for the \textbf{PUT} command in
  \figurename\ref{PUT_meas} are graphed here.  }
\label{SingleHop}
\end{figure}

\begin{figure}[!t]
\centering
\includegraphics[width=3in]{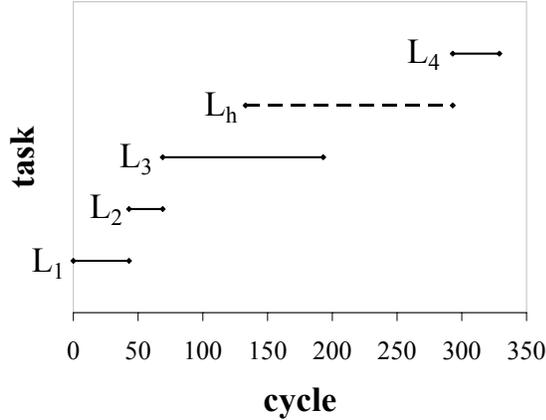}
\caption{Double-hop timings for a \textbf{PUT} command. $L_h$ is the
  time elapsed for the hop across an additional DNP, containing the
  trasmission time over the second serialized off-chip interface; it
  partially overlaps the $L_3$ for the first hop. }
\label{DoubleHop}
\end{figure}

Inter-tile on-chip ports are designed to be connected by
point-to-point parallel links and support a bandwidth of 1~word/cycle
as well. Because of the DNP distributed network architecture, it is
possible to arrange tiles inside the chip in a way to minimize the
wire lengths and the 
%corresponding capacitance effects on the
dissipated power due to data transmission.
In this ambit a parallel on-chip link is still a valid choice
considering the trade-off between power dissipation and bandwidth, as
the design of many NoC's demonstrates.

While in theory inter-tile off-chip ports may also sustain up to
1~word/cycle network load, the actual off-chip channel bandwidth is
inherently dependent on the serialization factor adopted --- defined
as the ratio between the DNP internal data width and the number of
serial lines --- on the link.
In the SHAPES implementation we chose a serialization factor equal to
16, in order to minimize the number of double data rate physical wires
(and corresponding multi-tile processor chip pin count), to allow for
high transmission frequencies and distances of the order of some
meters, and to simplify the design of the processing board PCB.
The above choice leads to a off-chip network bandwidth equal to 4
bit/cycle.
Taking into account all the factors, the inter-tile on-chip and
off-chip bandwidth are $BW_{on-chip} = N \times 32$ bit/cycle and
$BW_{off-chip} = M \times 4$ bit/cycle per direction.

%The associated single-hop latencies, defined as the time interval
%elapsed since the writing of the RDMA command on the sender DNP
%\emph{CMD FIFO} until the first word of the packet header is written
%on the destination DNP intra-tile interface (see
%\figurename\ref{PUT_meas}), are $L_{on-chip} = L_1+L_2+L_4 \sim 150$
%and $L_{off-chip} = L_1+L_2+L_3+L_4 \sim 250$ cycles, respectively
%300ns and 500ns at 500MHz. The cost in latency of additional hops over
%an off-chip interface, for example to reach the next-neighbor DNP,
%roughly cost another $L_2+L_3$ cycles.
The associated single-hop latencies (\figurename \ref{PUT_meas} and
\ref{SingleHop}), defined as the time interval elapsed since the
writing of the RDMA command on the sender DNP \emph{CMD FIFO} until
the first word of the packet header is written on the destination DNP
intra-tile interface, are $L_{on-chip} = L_1+L_2+L_4 \sim 130$ and
$L_{off-chip} = L_1+L_2+L_3+L_4 \sim 250$ cycles, respectively 260ns
and 500ns at 500MHz.
%
%The cost in latency of additional hops over an off-chip interface, for
%example to reach the next-neighbor DNP, is less than the expected
%$L_2+L_3$ cycles.
%
%This result is achieved by means of the wormhole routing algorithms
%
The cost in latency of an additional hop (\figurename\ref{DoubleHop})
over an off-chip interface, for example to reach the nextneighbor DNP,
is 100 cycles, which is less than the naive guess of $L2 + L3 \sim
150$ cycles thanks to wormhole routing.
By the way the relative high value of $l_{off-chip}$ is influenced by
the latency introduced by serialization and is dependent on the
serialization factor.

\begin{table}[!t]
%% increase table row spacing, adjust to taste
\renewcommand{\arraystretch}{1.3}
% if using array.sty, it might be a good idea to tweak the value of
% \extrarowheight as needed to properly center the text within the cells
\caption{Place\&Route trials for both MTNoC and MT2D DNP at 500MHz
  $45nm$ VLSI technology. Note that in both designs not all ports are
  used even though they are present and accounted for. MTNoC DNP area
  does not account for the additional on-chip NoC component.}
\label{tab_placeroute}
\centering
%% Some packages, such as MDW tools, offer better commands for making tables
%% than the plain LaTeX2e tabular which is used here.
\begin{tabular}{|c||c|c|}
\hline
                         & MTNoC DNP         & MT2D DNP           \\
\hline
\hline
on-chip ports (N)        & 1                 & 3                  \\
\hline
off-chip ports (M)       & 1                 & 1                  \\
\hline
estimated area           & 1.30$mm^{2}$      & 1.76$mm^{2}$       \\
\hline
estimated power          & 160mW             & 180mW              \\
\hline
\end{tabular}
\end{table}

We performed the Place\&Route trial of both the MTNoC and MT2D
inter-tile interconnection infrastructure, introduced in
sec. \ref{Exploring}, using 45nm silicon process at 500MHz.
The current DNP area estimates in the hypothesis of MTNoC and the MT2D
architectures are respectively $1.30 mm^2$ and $1.76 mm^2$; the larger
occupation area for the latter is mainly due to the higher number of
on-chip ports (3 ports vs. 1 port), implying a more complex switch
matrix architecture and a larger number of DNP data buffers, a
complexity which is moved in the NoC block in the MTNoC case.

%The estimated area occupation of the DNP appears to be rather large
%by current standards and not particularly outstanding;
Furthermore it must be taken into account that these estimates were
yielded by Place\&Route trials where the DNP data buffers were
synthesized using registers in place of memory macros, currently not
available in the silicon process library.  We expect to halve this
area in the final design.

Estimated figures for the power consumption for a single DNP in the
RDT computational tile are 160mW in the MTNoC case and 180mW in the
MT2D case, with the same silicon process and frequency, amounting
about to $1/4$ of the tile dissipation figure.

%The choice of using parallel channels was due to
%maximize bandwidth regardless of power consumption issues, as 
%Inter-tile off-chip bandwidth is strictly dependent on the serialization factor ...
%
%% More quantitative data about performance are currently in the
%% course of collection within the SHAPES project, as well as results
%% about power consumption, area and latency.
A motherboard built by 32 multi-tile processor chips (where one chip
includes 8 RDTs) would provide 1 Tera-Flops of floating point computing
power with roughly $600W$ of peak power consumption.

%------------------------------------------------------------------------------

\section{Conclusion and future work}

In the SHAPES project framework, we proved the parametric and
configurable architecture of the DNP to be effective for architectural
exploration of on-chip and off-chip interconnection networks, to
rapidly yield an efficient design.

As of today, the inter-tile off-chip interfaces are implemented with a
serialization factor equal to $16$; there is room for considerable
improvements in bandwidth and latency, either reducing the
serialization factor to $8$ or increasing the switching frequency of
the off-chip physical links. Using the target 45nm silicon process
technology we expect to double the current switching frequency pushing
it up to 1GHz.

The routing function of the network and the set of supported commands
are currently implemented by hard-coded logic blocks; the option to
instead have a $\mu$P in its place is currently under study. This
replacement would be a significant step ahead towards a more
thoroughly programmable DNP.

Moreover in future releases of the DNP we also plan to strengthen the
robustness of the network infrastructure integrating the minimal
hardware redundancy needed to support the well-known specific
fault-tolerant routing methods for torus-based, point-to-point network
~\cite{FAULT1}~\cite{FAULT2}.

Considering our know-how on massively parallel architectures,
collected in more than 20 years of research and development in the
field, we think that DNP can be a key feature to assemble HPC systems
that target multi-dimensional lattice problems of large proportions
(i.e.\ LQCD)~\cite{PETAPE}, for which high demands of parallel
processing and data communication between processing nodes are
considered critical.

%------------------------------------------------------------------------------

\section*{Acknowledgment}
The authors would like to thank Dr. R. Ammendola, for useful discussions and
vital activities on seminal work, and Prof. N. Cabibbo for being the
benign founder of the research group in which all of us do work.

%------------------------------------------------------------------------------


\IEEEtriggeratref{4}



%\bibliographystyle{IEEEtran}

\begin{thebibliography}{1}
\bibitem{Duato} J.~Duato, S.~Yalamanchili and L.~Ni,
  \emph{Interconnection Network - An engineering approach},\hskip 1em
  plus 0.5em minus 0.4em\relax Morgan Kaufmann Publishers Inc., San
  Francisco, CA, 2003.
\bibitem{Dally} W.J.~Dally and S.~Lacy, \emph{VLSI Architectures:
  Past, Present and Future}, \hskip 1em plus 0.5em minus 0.4em\relax
  Proc. Advanced Research in VLSI Conf., IEEE Press (1999) Pages
  232-241.
\bibitem{RAWtile} M.B. Taylor et al., \emph{The Raw Microprocessor: A
  Computational Fabric for Software Circuits and General-Pourpose
  Programs},\hskip 1em plus 0.5em minus 0.4em\relax IEEE Micro 2002.
\bibitem{IBMBGL} J.~Moreira et al., \emph{The Blue Gene/L
  Supercomputer: A Hardware and Software Story }, \hskip 1em plus
  0.5em minus 0.4em\relax International Journal of Parallel
  Programming, n.3, Pages 181-206,June 2007.
\bibitem{HDS} K.~Popovici, X.~Guerin, F.~Rousseau, P.S.~Paolucci and
  A.A.~Jerraya, \emph {Platform-based software design flow for
    heterogeneous MPSoC}, \hskip 1em plus 0.5em minus 0.4em\relax
  Transactions on Embedded Computing Systems, Vol.7, Pages 1-23, New
  York, NY, 2008, USA.
\bibitem{DOL} L.~Thiele, I.~Bacivarov, W.~Haid and K.~Huang,
  \emph{Mapping Applications to Tiled Multiprocessor Embedded
    Systems}, \hskip 1em plus 0.5em minus 0.4em\relax Proc. 7th Intl
  Conference on Application of Concurrency to System Design (ACSD
  2007), IEEE Computer Society, Pages 29-40, Bratislava, Slovak
  Republic, July 2007.
\bibitem{DNP1} P.S.~Paolucci, P.~Vicini et al., \emph{Introduction to
  the Tiled HW Architecture of SHAPES}, \hskip 1em plus 0.5em minus
  0.4em\relax Proc. Design, Automation and Test in Europe (DATE'07),
  Vol.1, Pages 77-82, France, Nice, April 2007.
\bibitem{SHAPES1} P.S.~Paolucci, A.A.~Jerraya, R.~Leupers, L.~Thiele,
  and P.~Vicini, \emph{Shapes: a tiled scalable software hardware
    architecture platform for embedded systems},\hskip 1em plus 0.5em
  minus 0.4em\relax Proceedings of the 4th international Conference on
  Hardware/Software Codesign and System Synthesis CODES+ISSS '06
  (Seoul, Korea), ACM Press, Pages 167-172, 2006.
\bibitem{DALLY2} W.J.~Dally and C.L.~Seitz, \emph{Deadlock-Free
  Message Routing in Multiprocessor Interconnection Networks}, \hskip
  1em plus 0.5em minus 0.4em\relax IEEE Trans. Comput. 36 (1987)
  547-553.
\bibitem{NoC1} M.~Coppola, M.~D. Grammatikakis, R.~Locatelli,
  G.~Maruccia and L.~Pieralisi, \emph{Design of Cost-Efficient
    Interconnect Processing Units: Spidergon STNoC}, \hskip 1em plus
  0.5em minus 0.4em\relax CRC press, Taylor \& Francis group, 2008.
\bibitem{NoC2} F.~Palumbo, S.~Secchi, D.~Pani, L.~Raffo, \emph{A Novel
  Non-Exclusive Dual-Mode Architecture for MPSoCs-Oriented Network on
  Chip Designs}, \hskip 1em plus 0.5em minus 0.4em\relax Proceedings
  of the International Workshop on Systems, Architectures, Modeling,
  and Simulation, Samos, Greece, July 21-24, 2008, Pages 96-105, LNCS
  5114.
\bibitem{NoC3} F.~Vitullo, N.~E. L'Insalata, E.~Petri, L.~Fanucci,
  M.~Casula, R.~Locatelli, M.~Coppola, \emph{Low-Complexity Link
    Microarchitecture for Mesochronous Communication in Networks on
    Chip}, \hskip 1em plus 0.5em minus 0.4em\relax IEEE Journal on
  Transactions on Computers, Vol. 57, No. 9, September 2008.
\bibitem{apeNEXT} F.~Bodin et al., \emph{The apeNEXT project}, \hskip
  1em plus 0.5em minus 0.4em\relax Nuclear Physics B - Proceedings
  Supplements, Volumes 106-107, March 2002, Pages 173-176, ISSN
  0920-5632.
\bibitem{APENET} R.~Ammendola et al., \emph{APENet: LQCD clusters a la
  APE},\hskip 1em plus 0.5em minus 0.4em\relax Nuclear Physics B
  (Proc. Suppl.) 140 (2005) 826-828
\bibitem{Huang08} K.~Huang, I.~Bacivarov, F.~Hugelshofer, and
  L.~Thiele, \emph{Scalably distributed systemc simulation for
    embedded applications}, \newblock In {\em Industrial Embedded
    Systems, 2008. SIES 2008.  International Symposium on}, Pages
  271--274, June 2008.
\bibitem{QCD} M.~Luscher, \emph{Lattice QCD: From quark confinement to
  asymptotic freedom}, \hskip 1em plus 0.5em minus 0.4em\relax Annales
  Henri Poincare 4:S197-S210,2003.
\bibitem{FAULT1} Rajendra V. Boppana and Suresh Chalasani,
  \emph{Fault-Tolerant Wormhole Routing Algorithms for Mesh
    Networks}, \hskip 1em plus 0.5em minus 0.4em\relax IEEE
  Transactions on computers, Vol. 44 n. 7, Pages 848-864, July 1995.
\bibitem{FAULT2} Rajendra V. Boppana and Suresh Chalasani,
  \emph{Fault-Tolerant Communication with Partitioned Dimension-Order
    Routers}, \hskip 1em plus 0.5em minus 0.4em\relax IEEE
  Transactions on Parallel and Distributed Systems, Vol. 10 n. 10,
  Pages 1026-1039, October 1999.
\bibitem{PETAPE} R.~Ammendola, A.~Biagioni, S.~De Luca, F.~Lo Cicero,
  A.~Lonardo, P.~S. Paolucci, M.~Perra, D.~Rossetti, C.~Sidore,
  F.~Simula, N.~Tantalo, L.~Tosoratto, P.~Vicini. \emph{Computing for
    Lattice QCD: New developments from the APE experiment}, \hskip 1em
  plus 0.5em minus 0.4em\relax Il Nuovo Cimento B, Vol. 123 n. 6-7,
  Pages 964-968 (2008).
%\bibitem{QCDOC} P.~Boyle et al., \emph{The QCDOC Project}, \hskip 1em
%  plus 0.5em minus 0.4em\relax Nuclear Physics B - Proceedings
%  Supplements, Vol. 140, LATTICE 2004 - Proceedings of the XXIInd
%  International Symposium on Lattice Field Theory, March 2005, Pages
%  169-175, ISSN 0920-5632.
%\bibitem{ANTON} D.E.~Shaw et al., \emph{Anton, a special-purpose
%  machine for molecular dynamics simulation}, \hskip 1em plus 0.5em
%  minus 0.4em\relax Proceedings of the 34th Annual international
%  Symposium on Computer Architecture (San Diego, California, USA, June
%  09 - 13, 2007). ISCA'07. ACM, New York, NY, 1-12.

\end{thebibliography}
%% \begin{thebibliography}{1}

%% \bibitem{IEEEhowto:kopka}
%% H.~Kopka and P.~W. Daly, \emph{A Guide to \LaTeX}, 3rd~ed.\hskip 1em plus
%%   0.5em minus 0.4em\relax Harlow, England: Addison-Wesley, 1999.

%% \end{thebibliography}

\end{document}